\documentclass[11pt, oneside]{article}   	
\usepackage[margin=1in]{geometry}               
\usepackage{graphicx}				
										
\usepackage{amssymb}
\usepackage{physics}
\usepackage{amsmath}
\usepackage[dvips]{epsfig}

\usepackage{amsfonts}
\usepackage{subfig}

\usepackage{mathrsfs}

\usepackage{xcolor}

\usepackage{bm}

\usepackage{blindtext}

\usepackage[title]{appendix}

\usepackage{cite}

\usepackage{authblk}

\def\cchi{\raise2pt\hbox{$\chi$}} 

\title{\bf{Some Comments on Unitary Qubit Lattice Algorithms for Classical Problems}}

\author{Paul Anderson ${ }^{1}$, Lillian Finegold-Sachs ${ }^{1}$, George Vahala ${ }^{1}$, Linda Vahala ${ }^{2}$, Abhay K.
Ram $^{3}$, Min Soe $^{4}$, Efstratios Koukoutsis ${ }^{5}$, Kyriakos Hizandis ${ }^{51}$\\

${ }^{1}$ Department of Physics, William \& Mary, Williamsburg, VA23185\\
${ }^{2}$ Department of Electrical \& Computer Engineering, Old Dominion University, Norfolk, VA\\
23529\\
${ }^{3}$ Plasma Science and Fusion Center, MIT, Cambridge, MA 02139\\
${ }^{4}$ Department of Mathematics and Physical Sciences, Rogers State University, Claremore,\\
OK 74017\\
${ }^{5}$ School of Electrical and Computer Engineering, National Technical University of Athens,\\
Zographou 15780, Greece}
\date{}

\begin{document}
\maketitle
\begin{abstract}
A qubit lattice algorithm (QLA), which consists of a set of interleaved unitary collision-streaming operators, is developed for electromagnetic wave propagation 
in tensor dielectric media.  External potential operators are required to handle gradients in the refractive indices, and these operators are typically non-unitary.
A similar problem arises in the QLA for the Korteweg-de Vries equation, as the potential operator  that models the KdV nonlinear term is also non-unitary.  Several QLAs are 
presented here that avoid the need of this non-unitary potential operator by perturbing the collision operator.  These QLAs are fully unitary.

\end{abstract}

\section{Introduction}

We have been investigating qubit lattice algorithms (QLA) for some time [1-21]. The aim of QLA is to develop a unitary interleaved sequence of collision-streaming operators which in the continuum limit reduces perturbatively to the desired differential equations describing the system of interest. The first step is to associate a basis set of qubits for the lattice, which on taking appropriate moments will recover the classical fields of interest.  Thus the QLA would be immediately encodable onto a quantum computer.  However, from our earlier nonlinear studies of 2D and 3D quantum turbulence [8,9, 11], the QLA is ideally parallelized on classical supercomputers with no degradation in parallel performance as the number of cores are ramped up [e.g., to over 750 000 cores on the $IBM$ BlueGene $Mira$ supercomputer at Argonne].   Some care is needed in the choice of the qubit basis.  For example, it will be shown in Sec. 2 that the simple basis choice of $(\mathbf{E,H})$ will never lead to a unitary representation of the Maxwell equations of electrodynamics. [$\mathbf{E}$ is the electric field, and $\mathbf{H}$ is the magnetic field].

At the heart of an efficient algorithm on a quantum computer is quantum entanglement of the qubits.  For example, consider a 2-qubit representation [e.g., for a basis for a  differential equation like the scalar nonlinear Schrodinger (NLS) equation or the Korteweg de Vries (KdV) equation].  A basis is the $2^{2}$  elements $(|00\rangle,|01\rangle,|10\rangle,|11\rangle)$.  Now consider a unitary $2 \times 2$ collision operator

\begin{equation}
C=\left[\begin{array}{cc}
\cos \theta & \sin \theta \\
-\sin \theta & \cos \theta
\end{array}\right]
\end{equation}

\noindent acting on the qubit subspace of $(|01\rangle,|10\rangle)$. One of the post-collision qubit elements is

\begin{equation}
\cos \theta \,|01\rangle+\sin \theta \,|10\rangle .
\end{equation}

\noindent However, this post-collision state cannot be represented by a tensor product of the $2^{2}$ - basis, since the most general tensor product state is

\begin{equation}
a_{0} b_{0}  |00\rangle + a_{0} b_{1}  |01\rangle + a_{1} b_{0} |10\rangle+a_{1} b_{1} |11\rangle
\end{equation}
where one qubit state is $a_{0} |0\rangle + a_{1}  |1\rangle$ , and the other state is $b_{0} |0\rangle + b_{1}  |1\rangle$ , 
\noindent for some coefficients $a_{0} \quad \ldots \quad b_{1}$.  To recover the state Eq. (2) from the tenor product state Eq. (3) one would  have to eliminate the $|00\rangle$-term.  However one must then set either $a_{0}=0$ or $b_{0}=0$.  But this would eliminate either the state $|01\rangle$ or the state $|10\rangle$ - both of which are needed to recover Eq. (2).  States which cannot be represented in a tensor product basis of qubits are called entangled states. A maximally entangled state is achieved on taking $\theta=\pi / 4$, and is known as a Bell state [22]

$$
B_{1}=\frac{|01\rangle+|10\rangle}{\sqrt{2}}
$$
\noindent  Note that the quantum entanglement is achieved here by the unitary collision operator.  The streaming operator in QLA will then propagate this entanglement throughtout the lattice.

In Sec. 2 we will develop a QLA for the solution of 2D Maxwell equations in a tensor Hermitian dielectric medium.  All our previous Maxwell QLA [18, 21] were restricted to scalar dielectrics.  We will present a simplified discussion of the Dyson map [23] that will permit us to transform from a non-unitary to unitary basis for the representation of 
the two curl equations of Maxwell.  From these qubit equations we will generate a QLA for tensor dielectric media that is second order accurate.  The QLA that we discuss here is not fully unitary.  While the collide-stream operator sequence is fully unitary, the external potential operators required to recover the Maxwell equations are not.  However these non-unitary matrices are very sparse and could be amenable to some unitary approximate representation.   
The role of the perturbation parameter $\epsilon$ introduced in the QLA is quite subtle.

As to an understanding of the subtlety of $\epsilon$ in QLA we return to the KdV equation in Sec. 3.  Our original QLA for KdV [1] consisted of maximally entanglement Bell collision operators together with a required external potential operator to model the KdV nonlinearity.  This external potential was not unitary.  Here we present a modified collision operator that is fully unitary and which leads to a QLA-KdV that does not require any external potential to be introduced.

Finally, in Sec. 4 we make some concluding remarks about future QLA simulations that are needed to elucidate the perturbation parameter $\epsilon$.  This parameter is required in order to move the discrete QLA into a continuum representation.

\section{QLA for Maxwell Equations}
\subsection{Qubit-Electromagnetic field representation}
Consider a simple dielectric non-magnetic medium with the constitutive equations

\begin{equation}
\mathbf{D}=\epsilon \mathbf{E}, \quad \mathbf{B}=\mu_{0} \mathbf{H} .
\end{equation}

Treating $\mathbf{u}=(\mathbf{E}, \mathbf{H})^{\mathbf{T}}$ as the fundamental fields, and $\mathbf{d}=(\mathbf{D}, \mathbf{B})^{\mathbf{T}}$ the derived fields , Eq. (2) can be written in matrix form

\begin{equation}
\mathbf{d}=\mathbf{W u}
\end{equation}

\noindent where $\mathbf{W}$ is a Hermitian $6 \times 6$ matrix

\begin{equation}
\mathbf{W}=\left[\begin{array}{cc}
\epsilon_{3 \times 3} & 0 \\
0 & \mu_{0} \mathbf{I}_{3 \times 3}
\end{array}\right]
\end{equation}

\noindent with $\mathbf{I}_{3 \times 3}$ the $3 \times 3$ identity matrix. and $\mathbf{T}$ is the transpose operator. The curl-curl (source-free) Maxwell equations $\nabla \times \mathbf{E}=-\partial \mathbf{B} / \partial t$, and $\nabla \times \mathbf{H}=\partial \mathbf{D} / \partial t$ in matrix form are just

\begin{equation}
i \frac{\partial \mathbf{d}}{\partial t}=\mathbf{M u}
\end{equation}

\noindent where, under standard boundary conditions, the curl-matrix operator $\mathbf{M}$ is Hermitian

\begin{equation}
\mathbf{M}=\left[\begin{array}{cc}
0_{3 \times 3} & i \nabla \times \\
-i \nabla \times & 0_{3 \times 3}
\end{array}\right]  .
\end{equation}

\noindent Since $\mathbf{W}$ is invertible, Eq. (5) can be written in terms of the basic electromagnetic fields $\mathbf{u}=(\mathbf{E}, \mathbf{H})$

\begin{equation}
i \frac{\partial \mathbf{u}}{\partial t}=\mathbf{W}^{-\mathbf{1}} \mathbf{M} \mathbf{u}
\end{equation}

In continuum applications, one typically treats the two Maxwell divergence equations $\nabla \cdot \mathbf{B}=0$ and $\nabla \cdot \mathbf{D}=0$ as initial conditions. From the curl-curl equations we see that they will then be satisfied for all time.

\subsubsection{homogeneous dielectric medium}
If one is dealing with a homogeneous dielectric medium (e.g., a vacuum), then the constitutive matrix $\mathbf{W}$ is a constant and trivially commutes with the curl-operator $\mathbf{M}$. As a result, the product of the two Hermitian matrices, $\mathbf{W}^{-\mathbf{1}} \mathbf{M}$ is itself Hermitian, and Eq. (7) gives a unitary evolution of the electromagnetic fields $\mathbf{u}=(\mathbf{E}, \mathbf{H})^{\mathbf{T}}$.  Thus $\mathbf{u}$ is an appropriate basis for the qubit fields and for quantum computation.

\subsubsection{inhomogeneous dielectric media}
However, when the matrix $\mathbf{W}$ is spatiaily dependent, then $\mathbf{W}^{-\mathbf{1}} \mathbf{M} \neq \mathbf{M} \mathbf{W}^{-\mathbf{1}}$ and $\mathbf{W}^{-\mathbf{1}} \mathbf{M}$ is not Hermitian. Under these conditions, the qubit representation of the electromagnetic fields $\mathbf{u}=(\mathbf{E}, \mathbf{H})^{\mathbf{T}}$ will not yield a unitary evolution of these qubits. However Koukoutsis et. al. [23] have shown how to determine the so-called Dyson map from the fields $\mathbf{u}$ to a new field representation $\mathbf{U}$ such that the resultant representation in terms of the new field $\mathbf{U}$ will result in a unitary evolution of these fields. Indeed, it can be shown [23], that the Dyson map

\begin{equation}
\mathrm{U}=\mathrm{W}^{1 / 2} \mathrm{u}
\end{equation}

\noindent will yield a unitary evolution equation for $\mathbf{U}$ with

\begin{equation}
i \frac{\partial \mathbf{U}}{\partial t}=\mathbf{W}^{-1 / 2} \mathbf{M} \mathbf{W}^{-1 / 2} \mathbf{U}
\end{equation}

\noindent as the matrix operator $\mathbf{W}^{-\mathbf{1} / \mathbf{2}} \mathbf{M} \mathbf{W}^{-\mathbf{1} / \mathbf{2}}$ is Hermitian.

Thus one could start to build a QLA based on the electromagnetic fields

\begin{equation}
\mathbf{U}=\left(\epsilon^{1 / 2} \mathbf{E}, \mu_{0}^{1 / 2} \mathbf{H}\right)^{T}
\end{equation}

\noindent or under the rotation matrix

\begin{equation}
\mathbf{L}=\frac{1}{\sqrt{2}}\left[\begin{array}{cc}
I_{3 \times 3} & i I_{3 \times 3} \\
I_{3 \times 3} & -i I_{3 \times 3}
\end{array}\right]
\end{equation}

\noindent one could base a QLA on the field representation $\mathbf{U}_{\mathbf{R S W}}=\mathbf{L U}$ where

\begin{equation}
\mathbf{U}_{\mathbf{R S W}}=\frac{1}{\sqrt{2}}\left[\begin{array}{c}
\epsilon^{1 / 2} \mathbf{E}+i \mu_{0}^{1 / 2} \mathbf{H} \\
\epsilon^{1 / 2} \mathbf{E}-i \mu_{0}^{1 / 2} \mathbf{H}
\end{array}\right] .
\end{equation}

\noindent This is nothing but the unitary evolution of the Riemann-Silberstein-Weber (RSW) vector - a representation used to represent Maxwell equations from the early 1920's [24-26].  

Moreover, the theory can be readily extended to diagonal tensor dielectric media, with (assuming non-magnetic materials) the 6-qubit representation $\mathbf{Q}$ of the field
\begin{equation}
\mathbf{U}=\left(n_x E_x, n_y E_y, n_z E_z, \mu_{0}^{1 / 2} \mathbf{H}\right)^{T}  = \mathbf{Q}
\end{equation}
\noindent $(n_x , n_y, n_z)$ is the vector (diagonal) refractive index, with $\epsilon_x = n_x^2$ ...  .  We work in Cartesian coordinates.

\subsection{$ \mathrm{2D}$ QLA for $\mathrm{x-y}$ dependent propagation of Maxwell Equations}
From Eqs. (9) and (13), Maxwell equations for 2D x-y spatially dependent fields written in terms of the 6-$\mathbf{Q}$ vector components
\begin{equation}
\begin{aligned}
\frac{\partial q_0}{\partial t} = \frac{1}{n_x} \frac{\partial q_5}{\partial y} , \qquad
\frac{\partial q_1}{\partial t} = - \frac{1}{n_y} \frac{\partial q_5}{\partial x} , \qquad
\frac{\partial q_2}{\partial t} =  \frac{1}{n_z} \left[ \frac{\partial q_4}{\partial x} -\frac{\partial q_3}{\partial y} \right] \\
\frac{\partial q_3}{\partial t} = - \frac{\partial (q_2/n_z)}{\partial y} , \qquad
\frac{\partial q_4}{\partial t} = \frac{\partial (q_2/n_z)}{\partial x} , \qquad
\frac{\partial q_5}{\partial t} = - \frac{\partial (q_1/n_y)}{\partial x}  + \frac{\partial (q_0/n_x)}{\partial y} 
\end{aligned}
\end{equation}
\noindent This representation is unitary.

Our QLA representation focusses on recovering Eq. (14) perturbatively.  One can thus consider developing the representation dimension by dimension.  In particular
we introduce the following unitary collision operator  with collision angles $\theta_1$ and $\theta_2$ (to be specified later):
\begin{equation}
C_X=\left[\begin{array}{cccccc}
1 & 0 & 0 & 0& 0& 0 \\
0 & cos \,\theta_1 & 0 & 0 & 0 & - sin\,\theta_1 \\
0 & 0 & cos\,  \theta_2 & 0 & - sin \,\theta_2 & 0 \\
0 & 0 & 0 & 1 &  0 & 0  \\
0 & 0 & sin\,\theta_2 & 0 & cos\, \theta_2 & 0 \\
0 & sin\, \theta_1 & 0 & 0 & 0 & cos \,\theta_1
\end{array}\right]
\end{equation}
\noindent and the unitary collision operator
\begin{equation}
C_Y=\left[\begin{array}{cccccc}
cos\, \theta_0 & 0 & 0 & 0& 0& sin\, \theta_0 \\
0 & 1 & 0 & 0 & 0 & 0 \\
0 & 0 & cos\,  \theta_2 & 0 &  sin \,\theta_2 & 0 \\
0 & 0 & -sin\,\theta_2 & cos \,\theta_2 &  0 & 0  \\
0 & 0 & 0& 0 & 1 & 0 \\
-sin \,\theta_0& 0 & 0 & 0 & 0 & cos \,\theta_0
\end{array}\right]
\end{equation}
\noindent with collision angles $\theta_0$ and $\theta_2$.  The unitary streaming operator $S^{+x}_{14}$ shifts qubits $q_1$ and $q_4$ one lattice unit in the $+x$ direction, while leaving the remaining 4 qubits alone.  We finally need to introduce the external potential operators
\begin{equation}
V_X=\left[\begin{array}{cccccc}
1 & 0 & 0 & 0& 0& 0 \\
0 & 1 & 0 & 0 & 0 & 0\\
0 & 0 & 1 & 0 &0& 0 \\
0 & 0 & 0 & 1 &  0 & 0  \\
0 & 0 &- sin\,\beta_2 & 0 & cos\, \beta_2 & 0 \\
0 & sin\, \beta_0 & 0 & 0 & 0 & cos \,\beta_0
\end{array}\right]
\end{equation}
\noindent and
\begin{equation}
V_Y=\left[\begin{array}{cccccc}
1 & 0 & 0 & 0& 0& o \\
0 & 1 & 0 & 0 & 0 & 0\\
0 & 0 & 1 & 0 &0& 0 \\
0 & 0 & \cos\, \beta_3& \sin \, \beta_3 &  0 & 0  \\
0 & 0 & 0 & 0 & 1 & 0 \\
-sin\, \beta_1 & 0 & 0 & 0 & 0 & cos \,\beta_1
\end{array}\right]
\end{equation}
\noindent  for particular angles $\beta_0$, $\beta_1$ and $\beta_2$.  These potential operators are not unitary, but very sparse.

We now consider the following unitary sequence of interleaved collision-streaming operators:
\begin{equation}
\mathbf{U_X} = S^{+x}_{25}.C_X^\dag . S^{-x}_{25}.C_X. S^{-x}_{14}.C_X^\dag . S^{+x}_{14}.C_X .S^{-x}_{25}.C_X . S^{+x}_{25}.C_X^\dag. S^{+x}_{14}.C_X . S^{-x}_{14}.C_X^\dag 
\end{equation}
\noindent  and
\begin{equation}
\mathbf{U_Y} = S^{+y}_{25}.C_Y^\dag . S^{-y}_{25}.C_Y. S^{-y}_{03}.C_Y^\dag . S^{+y}_{03}.C_Y .S^{-y}_{25}.C_Y . S^{+y}_{25}.C_Y^\dag. S^{+y}_{03}.C_Y . S^{-y}_{03}.C_Y^\dag 
\end{equation}
\noindent with the discrete time advancement of the 6-qubit $\mathbf{Q}$ given by
\begin{equation}
\mathbf{Q}(t+\delta t) = V_Y.V_X.\mathbf{U_Y}. \mathbf{U_X}. \mathbf{Q}(t)
\end{equation}

To recover the desired Maxwell equations (14) perturbatively, one employ a small parameter $\epsilon$ as the spatial lattice shift unit (assuming a square $x-y$ lattice),
and the unitary collision angles
\begin{equation}
\theta_0 = \frac{\epsilon}{4 n_x} \quad , \qquad \theta_1 = \frac{\epsilon}{4 n_y} \quad , \qquad  \theta_2 = \frac{\epsilon}{4 n_z} .
\end{equation}
so as to recover the coefficients of the $\partial \mathbf{Q}/\partial (x,y)$ terms.
\noindent Finally, the nonunitary external potential angles need to be defined as
\begin{equation}
\beta_0 = \epsilon^2 \frac{\partial n_y/\partial x}{n^2_y} \quad , \quad \beta_1 = \epsilon^2 \frac{\partial n_x/\partial y}{n^2_x}  \quad , \quad \beta_2 = \epsilon^2 \frac{\partial n_z/\partial x}{n^2_z} \quad , \quad \beta_3 = \epsilon^2 \frac{\partial n_z/\partial y}{n^2_z}
\end{equation}
Indeed, using Mathematica to evaluate Eq. (21), one obtains in the continuum spatial limit the desired Maxwell equations to errors of $\epsilon^4$
\begin{equation}
\begin{aligned}
\frac{\partial q_0}{\partial t} = \epsilon^2 \delta t \frac{1}{n_x} \frac{\partial q_5}{\partial y} , \qquad
\frac{\partial q_1}{\partial t} = - \epsilon^2  \delta t  \frac{1}{n_y} \frac{\partial q_5}{\partial x} , \qquad
\frac{\partial q_2}{\partial t} =  \epsilon^2 \delta t   \frac{1}{n_z} \left[ \frac{\partial q_4}{\partial x} -\frac{\partial q_3}{\partial y} \right] \\
\frac{\partial q_3}{\partial t} = - \epsilon^2 \delta t   \frac{\partial (q_2/n_z)}{\partial y} , \qquad
\frac{\partial q_4}{\partial t} = \epsilon^2 \delta t   \frac{\partial (q_2/n_z)}{\partial x} , \qquad
\frac{\partial q_5}{\partial t} = -\epsilon^2 \delta t   \left( \frac{\partial (q_1/n_y)}{\partial x}  + \frac{\partial (q_0/n_x)}{\partial y} \right)
\end{aligned}
\end{equation}

\noindent i.e., under diffusion ordering, $\epsilon^2 \delta t \approx O(1)$, one recovers the continuum Maxwell equations to errors $O(\epsilon^2)$.  We will explore QLA simulations of 2D Maxwell equations in a subsequent paper.

\section{QLA for KdV without external non-unitary potential operators}
The KdV equation is an important nonlinear equation and was developed to explore the evolution of shallow water waves. Interestingly, it [27] has also been associated with the Fermi-Pasta-Ulam-Tsingou simulations of the 1950's.  Fermi wanted to examine the equipartition of energy among the modes of a many body problem of weakly coupled nonlinear oscillators. 
Statistical mechanics indicates that the time- asymptotic state will be one in which there is equipartition of energy among all the oscillator modes.
Instead, in the parameter regime they considered, Fermi et. al. found recurrence of initial conditions, but a recurrence that was not a usually extremely long Poincare recurrence time of Hamiltonian systems.  Interestingly, this would turn out to be a precursor to soliton theory.

 The general $\mathrm{KdV}$ equation for arbitrary positive constants $a$ and $b$

\begin{equation}
\frac{\partial \psi}{\partial t}+a \psi \frac{\partial \psi}{\partial x}+b \frac{\partial^{3} \psi}{\partial x^{3}}=0
\end{equation}

\noindent is exactly integrable. One of its solutions is the right traveling soliton with speed $c$ - a free parameter

\begin{equation}
\psi(x, t)=\frac{3 c}{a} \operatorname{sech}^{2}\left(\frac{1}{2} \sqrt{\frac{c}{b}}[x-c t]\right)
\end{equation}
Notice that for the KdV soliton, the amplitude and its speed are correlated (unlike the NLS soliton).

 Since the $\mathrm{KdV}$ equation is a scalar equation for the real function $\psi(x, t)$ one need only to employ 2 qubits / lattice site. First we shall reconsider the QLA for KdV with the use of an external potential to model the nonlinear term in KdV [1]. The collision operator is nothing but Eq. (1).  We denote the operator $S_0^+$ to be the streaming operator that translates the qubit $q_{0}$ one lattice unit in the $+x$-direction. To eliminate the 2nd order spatial derivative one must choose the interleaved sequence of collision-stream unitary operators carefully. In particular the following sequence will generate a second order QLA for the KdV equation

\begin{equation}
Q(t+\Delta t)=V_{pot} \cdot S_0^+ C \cdot S_1^- C^{T} \cdot S_0^- C \cdot S_1^+ C^{T} \cdot S_0^- C^{T} \cdot S_1^+ C \cdot S_0^+ C^{T} \cdot S_1^- C \cdot Q(t)
\end{equation}

\noindent where the unitary collision operator $C$ is nothing but the maximally entangling operator. Eq, (1), with $\theta = \pi/4$. $Q=\left(q_{0} \, q_{1}\right)^{T}$. The external potential $V_{\text {pot }}$ is the Hermitian matrix

\begin{equation}
V_{\text {pot }}=\left[\begin{array}{cc}
\cos \alpha & -\sin \alpha \\
-\sin \alpha & \cos \alpha
\end{array}\right] \quad \text { with } \quad \alpha=\epsilon^{3} m[x] .
\end{equation}

In the continuum limit, one recovers

\begin{equation}
\frac{\partial \psi}{\partial t}+\epsilon^{3}\left(m[x] \cdot \psi(x, t)+\frac{1}{2} \frac{\partial^{3} \psi}{\partial x^{3}}\right)=0+O\left(\epsilon^{5}\right)
\end{equation}

\noindent on defining $\psi=q_{0}+q_{1}$. With the choice of $m[x]=\partial \psi / \partial x$ we have a second order accurate QLA for KdV. Note that the QLA of Eq. (26) is not fully unitary because of the non-unitary property of the external potential operator $V_{\text {pot }}$.

\subsection{Fully unitary QLAs for KdV}
There is a large class of unitary QLAs all of which recover the KdV equation to second order accuracy.
 Here, we will present two QLAs, both having the same unitary collision operator, but with different streaming sequences on the two qubits. Indeed, using Mathematica, it can be shown that the following QLA

\begin{equation}
Q(t+\Delta t)=S_0^- C_{1} \cdot S_0^+ C_{1} \cdot S_1^+ C_{1} \cdot S_1^- C_{1}^{T} \cdot S_0^- C_{1}^{T} \cdot S_0^+ C_{1}^{T} \cdot S_1^+ C_{1}^{T} \cdot S_1^- C_{1} \cdot Q(t)
\end{equation}

\noindent with unitary collision operator $C_{1}$

\begin{equation}
C_{1}=\left[\begin{array}{cc}
\cos \alpha_{1} & \sin \alpha_{1} \\
-\sin \alpha_{1} & \cos \alpha_{1}
\end{array}\right] \quad \text { with } \quad \alpha_{1}=\frac{\pi}{4} +   \epsilon^{2} m_{1}[x] .
\end{equation}

\noindent leads in the continuum limit to

\begin{equation}
\frac{\partial \psi_{1}}{\partial t}+\epsilon^{3}\left(4 m_1[x] \cdot \frac{\partial \psi_{1}}{\partial x}+\frac{1}{2} \frac{\partial^{3} \psi_{1}}{\partial x^{3}}\right)=0+O\left(\epsilon^{5}\right)
\end{equation}

\noindent so that the choice of $m[x]=\psi_{1}$ will recover $\mathrm{KdV}$.

Another fully unitary QLA that recovers KdV has the following interleaved sequence of unitary collision-streaming operators:

\begin{equation}
Q(t+\Delta t)=C_{1} S_0^-\cdot C_{1} S_1^+ \cdot C_{1} S_0^- \cdot C_{1} S_1^+ \cdot C_{1}^{T} S_0^+ \cdot C_{1}^{T} S_1^- \cdot C_{1}^{T} S_0^+ \cdot C_{1}^{T} S_1^- \cdot Q(t)
\end{equation}

\noindent $C_{1}$ is the same collision operator, Eq. (33).  In the continuum limit, we find

\begin{equation}
\frac{\partial \psi_{1}}{\partial t}+\epsilon^{3}\left(-4 m_1[x] \cdot \frac{\partial \psi_{1}}{\partial x}+\frac{1}{2} \frac{\partial^{3} \psi_{1}}{\partial x^{3}}\right)=0+O\left(\epsilon^{5}\right)
\end{equation}

\noindent so that the choice of $m_1[x] = - \psi_1$ will recover the KdV equation.

The implementation of these fully unitary algorithms may not necessarily be straightforward as the perturbation parameter $\epsilon$ introduced into the Mathematica algorithm requires a perturbation in the collision angle of $O\left(\epsilon^{2}\right)$, Eq. (33), while the continuum limit has scaling proceeds as $O\left(\epsilon^{3}\right)$. In previous QLA for nonlinear physics, the order of the function $\psi$ controlled the $\epsilon$-factor.

\section{SUMMARY}
The development of a fully unitary QLA for plasma physics [28-31] in particular,  is of considerable interest to us as these algorithms
 are immediately encodable on quantum computers. Of course, since they are time evolution algorithms, they will have to wait till there are error-correcting quantum
 computers available.  
 In developing QLAs for plasma physics we have taken the tack of first concentrating on the Maxwell equations in a given scalar dielectric media.  Then one would eventually generalize to a tensor unitary dielectric description of a cold magnetized plasma.
 
 Here, we have shown how to generalize our QLA-scalar dielectric Maxwell equations to handle tensor Hermitian dielectric media.  This was facilitated by the use of the Dyson map [23].  Indeed, the explicit determination of the Dyson map  proves that there exists a unitary quantum algorithm to describe such a Maxwell system.  The problem, of course, is to explicitly construct such an algorithm.  We have concentrated on the QLA approach, which employs a
 non-trivial sequence of interleaved collide-stream unitary operators. To proceed explicitly with QLA, one must resort to perturbation theory and the introduction of a small parameter $\epsilon$.   The uinitary collide-stream operator sequence does not in all our Maxwell equation considerations recover the required full set of evolution equations.
 This has resulted in the need to introduce so-called potential operators in order to recover the equation of interest.  At least one of these potential operators turns out to be non-unitary.  As these QLAs have parallelized outstandingly on classical supercomputer architectures, outperforming standard computational fluid dynamic codes for the study of quantum turbulence, we have proceeded with the numerical implementation of such QLAs.  This seems prudent as an error-correcting quantum computer with
 long qubit coherence times is still on the somewhat distant horizon.  Nevertheless we are also pursuing a fully unitary QLA.  
 In particular, we are revisiting our original non-unitary QLA-KdV [1] to determine a fully unitary QLA.  We have found a large class of such unitary QLA-KdV, based on 
 changing the particular collide-stream sequences.  
 The implementation of these fully unitary algorithms may not necessarily be straightforward as the perturbation parameter $\epsilon$ introduced into the Mathematica symbolic manipulations requires a perturbation in the collision angle of $O\left(\epsilon^{2}\right)$, Eq. (33), while the continuum limit has scaling proceeds as $O\left(\epsilon^{3}\right)$, Eq. (34) or (36).   In previous QLA for nonlinear physics, the order of the function $\psi$ was used as the $\epsilon$-factor.  
 We believe that understanding the role of $\epsilon$ in the QLA-KdV simulations will be pivotal in handling the role of $\epsilon$ in QLA-Maxwell in both scalar and tensor dielectric media.  These simulations will be reported in a future publication.

\section{Acknowledgments}
This research was partially supported by Department of Energy grants DE-SC0021647, DE-FG0291ER-54109, DE-SC0021651, DE-SC0021857, and DE-SC0021653. This work has been carried out partially within the framework of the EUROfusion Consortium. E.K has received funding from the Euratom research and training program WPEDU under grant agreement no. 101052200 as well as from the National Program for Controlled Thermonuclear Fusion, Hellenic Republic. K.H is
supported by the National Program for Controlled Thermonuclear Fusion, Hellenic Republic. The views and opinions expressed herein do not necessarily reflect those of the European Commission.

\section{References}
\qquad [1] VAHALA, G, VAHALA, L \& YEPEZ, J. 2003 Quantum lattice gas representation of some classical solitons. Phys. Lett A310, 187-196

[2] VAHALA, L, VAHALA, G \& YEPEZ, J. 2003 Lattice Boltzmann and quantum lattice gas representations of one-dimensional magnetohydrodynamic turbulence. Phys. Lett A306, 227-234.

[3] VAHALA, G, VAHALA, L \& YEPEZ, J. 2004. Inelastic vector soliton collisions: a latticebased quantum representation. Phil. Trans: Mathematical, Physical and Engineering Sciences, The Royal Society, 362, 1677-1690 [4] VAHALA, G, VAHALA, L \& YEPEZ, J. 2005 Quantum lattice representations for vector solitons in external potentials. Physica A362, 215-221.

[5] YEPEZ, J. 2002 An efficient and accurate quantum algorithm for the Dirac equation. arXiv: 0210093.

[6] YEPEZ, J. 2005 Relativistic Path Integral as a Lattice-Based Quantum Algorithm. Quant. Info. Proc. 4, 471-509.

[7] YEPEZ, J, VAHALA, G \& VAHALA, L. 2009a Vortex-antivortex pair in a Bose-Einstein condensate, Quantum lattice gas model of theory in the mean-field approximation. Euro. Phys. J. Special Topics 171, 9-14

[8] YEPEZ, J, VAHALA, G, VAHALA, L \& SOE, M. 2009b Superfluid turbulence from quantum Kelvin wave to classical Kolmogorov cascades. Phys. Rev. Lett. 103, 084501.

[9] VAHALA, G, YEPEZ, J, VAHALA, L, SOE, M, ZHANG, B, \& ZIEGELER, S. 2011 Poincaré recurrence and spectral cascades in three-dimensional quantum turbulence. Phys. Rev. E84, 046713

[10] VAHALA, G, YEPEZ, J, VAHALA, L \&SOE, M, 2012 Unitary qubit lattice simulations of complex vortex structures. Comput. Sci. Discovery 5, 014013

[11] VAHALA, G, ZHANG, B, YEPEZ, J, VAHALA. L \& SOE, M. 2012 Unitary Qubit Lattice Gas Representation of 2D and 3D Quantum Turbulence. Chpt. 11 (pp. 239 - 272), in Advanced Fluid Dynamics, ed. H. W. Oh, (InTech Publishers, Croatia)

[12] YEPEZ, J. 2016 Quantum lattice gas algorithmic representation of gauge field theory. SPIE 9996, paper 9996-2

[13] OGANESOV, A, VAHALA, G, VAHALA, L, YEPEZ, J \& SOE, M. 2016a. Benchmarking the Dirac-generated unitary lattice qubit collision-stream algorithm for 1D vector Manakov soliton collisions. Computers Math. with Applic. 72, 386

[14] OGANESOV, A, FLINT, C, VAHALA, G, VAHALA, L, YEPEZ, J \& SOE, M 2016b Imaginary time integration method using a quantum lattice gas approach. Rad Effects Defects Solids $171,96-102$

[15] OGANESOV, A, VAHALA, G, VAHALA, L \& SOE, M. 2018. Effects of Fourier Transform on the streaming in quantum lattice gas algorithms. Rad. Eff. Def. Solids, 173, 169-174

[16] VAHALA, G., SOE, M., VAHALA, L., \& RAM, A. K., 2021 One- and Two-Dimensional quantum lattice algorithms for Maxwell equations in inhomogeneous scalar dielectric media I : theory. Rad. Eff. Def. Solids 176, 49-63.

[17] VAHALA, G., SOE, M., VAHALA, L., \& RAM, A. K., 2021 One- and Two-Dimensional quantum lattice algorithms for Maxwell equations in inhomogeneous scalar dielectric media II : Simulations. Rad. Eff. Def. Solids 176, 64-72.

[18] VAHALA, G, VAHALA, L, SOE, M \& RAM, A, K. 2020. Unitary Quantum Lattice Simulations for Maxwell Equations in Vacuum and in Dielectric Media, J. Plasma Phys 86, 905860518

[19] VAHALA, L, SOE, M, VAHALA, G \& YEPEZ, J. 2019a. Unitary qubit lattice algorithms for spin-1 Bose-Einstein condensates. Rad Eff. Def. Solids 174, 46-55

[20] VAHALA, L, VAHALA, G, SOE, M, RAM, A \& YEPEZ, J. 2019b. Unitary qubit lattice algorithm for three-dimensional vortex solitons in hyperbolic self-defocusing media. Commun Nonlinear Sci Numer Simulat 75, 152-159

[21] RAM, A. K., VAHALA, G., VAHALA, L. \& SOE, M 2021 Reflection and transmission of electromagnetic pulses at a planar dielectric interface - theory and quantum lattice simulations AIP Advances 11, 105116 (1-12). 

[22] MERMIN, N. D., 2007 Quantum computer science, Cambridge University Press, Cambridge 

[23] KOUKOUTSIS, E., HIZANIDIS, K., RAM, A. K., \& VAHALA, G. 2022. Dyson Maps and Unitary Evolution for Maxwell Equations in Tensor Dielectric Media. arXiv:2209.08523

[24] LAPORTE, O. \& UHLENBECK, G. E. 1931 Application of spinor analysis to the Maxwell and Dirac equations. Phys. Rev. 37, 1380-1397.

[25] OPPENHEIMER, J. R. 1931 Note on light quanta and the electromagnetic field. Phys. Rev. 38, 725-746.

[26] MOSES, E. 1959 Solutions of Maxwell's equations in terms of a spinor notation: the direct and inverse problems, Phys. Rev. 113, 1670-1679

[27] BERMAN, FG. P., \& IZRAILEV, F. M., 2005 The Fermi-Pasta-Ulam problem: Fifty years of progress. Chaos 15, 015104.

[28] DODIN, I. Y. \& STARTEV, E. A. 2021 On appliations of quantum computing to plasma simulations. Phys. Plasmas 28, 092101

[29] ENGEL, A., SMITH, G \& PARKER, S. E. 2019 Quantum Algorithm for the Vlasov Equation, Phys. Rev. 100, 062315

[30] ENGEL, A., SMITH, G \& PARKER, S. E. 2021. Linear Embedding of nonlinear dynamial systems and prospects for efficient quantum algorithms, Phys. Plasmas 28, 062305

[31] LIU, J-P, KOLDEN, H.O, KROVI, H. K, LOUREIRO, N. F, TRIVISA, K, \& CHILDS, A. M. 2021. Proc. Natl. Acad. Sciences 118, e2026805118.

\end{document}